\documentclass[conference]{IEEEtran}
\usepackage{epsfig}
\usepackage{times}
\usepackage{float}
\usepackage{afterpage}
\usepackage{amsmath}
\usepackage{amstext}
\usepackage{amssymb,bm}
\usepackage{latexsym}
\usepackage{color}
\usepackage{graphicx}
\usepackage{amsmath}
\usepackage{amsthm}
\usepackage{graphicx}
\usepackage[center]{caption}
\usepackage{pstricks}
\usepackage{caption}
\usepackage{subcaption}
\usepackage{booktabs}
\usepackage{multicol}
\usepackage{lipsum}
 \usepackage[T1]{fontenc}
\usepackage{epstopdf}

\allowdisplaybreaks

\newtheorem{thm}{Theorem}

\newtheorem{rem}{Remark}

\theoremstyle{definition}

\providecommand{\definitionname}{Definition}

\newfloat{algorithm}{tbp}{loa}
\providecommand{\algorithmname}{Algorithm}
\floatname{algorithm}{\protect\algorithmname}

\usepackage{changes}
\definechangesauthor[color=blue,name={Mingyue Ji}]{MJ}

\long\def\comment#1{}

\newcommand{\dv}{{\mathbf d}}

\newcommand{\Zm}{{\mathbf Z}}

\newcommand{\Ac}{{\mathcal A}}
\newcommand{\Bc}{{\mathcal B}}
\newcommand{\Cc}{{\mathcal C}}

\newcommand{\Hc}{{\mathcal H}}

\newcommand{\Jc}{{\mathcal J}}

\newcommand{\Lc}{{\mathcal L}}

\newcommand{\Pc}{{\mathcal P}}
\newcommand{\Qc}{{\mathcal Q}}

\newcommand{\Sc}{{\mathcal S}}

\newcommand{\Uc}{{\mathcal U}}
\newcommand{\Wc}{{\mathcal W}}
\newcommand{\Vc}{{\mathcal V}}
\newcommand{\Xc}{{\mathcal X}}

\newcommand{\fsf}{{\mathsf f}}

\newcommand{\rsf}{{\mathsf r}}

\newcommand{\Bsf}{{\mathsf B}}

\newcommand{\Hsf}{{\mathsf H}}

\newcommand{\Ksf}{{\mathsf K}}

\newcommand{\Msf}{{\mathsf M}}
\newcommand{\Nsf}{{\mathsf N}}

\newcommand{\Rsf}{{\mathsf R}}

\renewcommand{\arg}{{\hbox{arg}}}

\newcommand{\be}{\begin{equation}}
\newcommand{\ee}{\end{equation}}
\newcommand{\bea}{\begin{eqnarray}}
\newcommand{\eea}{\end{eqnarray}}

\def\fsf{ {\mathsf f}}

\begin{document}

\title{Caching in Combination Networks: A Novel Delivery  by Leveraging the Network Topology}

\author{
\IEEEauthorblockN{%
Kai~Wan\IEEEauthorrefmark{1}, %
Mingyue~Ji\IEEEauthorrefmark{2}, %
Pablo~Piantanida\IEEEauthorrefmark{1}, %
Daniela~Tuninetti\IEEEauthorrefmark{3}%
}
\IEEEauthorblockA{\IEEEauthorrefmark{1}L2S CentraleSupélec-CNRS-Université Paris-Sud, 
France, \{kai.wan, pablo.piantanida\}@l2s.centralesupelec.fr}%
\IEEEauthorblockA{\IEEEauthorrefmark{2}University of Utah, Salt Lake City, 
USA,  mingyue.ji@utah.edu}%
\IEEEauthorblockA{\IEEEauthorrefmark{3}University of Illinois at Chicago, Chicago, 
USA, danielat@uic.edu}%
}

\maketitle

\begin{abstract}
\emph{THIS PAPER IS ELIGIBLE FOR THE STUDENT PAPER AWARD.} 
Maddah-Ali and Niesen (MAN) in 2014 surprisingly showed that it is possible to serve an arbitrarily large number of cache-equipped users with a constant number of transmissions by using coded caching 
in shared-link broadcast networks. 
This paper studies the tradeoff between the user's cache size and the file download time for {\it combination networks}, where users with caches communicate with the servers through intermediate relays.
%
Motivated by the so-called separation approach, it is assumed that placement and multicast message generation are done according to the MAN original scheme and regardless of the network topology.
The main contribution of this paper is the design of a novel two-phase delivery scheme that, accounting to the network topology, outperforms schemes available in the literature.
The key 
idea is to create additional (compared to MAN) multicasting opportunities:
in the first phase coded messages are sent with the goal of increasing the amount of `side information' at the users, which is then leveraged during the second phase.
The download time with the novel scheme is shown to be proportional to $1/\Hsf$ (with $\Hsf$ being the number or relays) and to be order optimal under the constraint of uncoded placement for some parameter regimes. 
\end{abstract}

\section{Introduction}
\label{sec:intro}

Network traffic can be smoothed out to some extent
by placing content in user local memories during off-peak hours ({\it placement phase}) 
with the hope that the pre-fetched content will be requested during peak hours, in which case the number of broadcast transmissions from the server to the users ({\it delivery phase}) will be reduced.
Coded caching was originally considered in~\cite{dvbt2fundamental} by Maddah-Ali and Niesen (MAN) for 
 shared-link broadcast networks, where a server (equipped with $\Nsf$ files) communicates to $\Ksf$ users (with a cache able to store $\Msf$ files) through a shared noiseless channel.  
In the original MAN scheme, each of the $\Nsf$ files is partitioned into a number of pieces, and each piece is placed uncoded into a number of user caches that depends on the cache size $\Msf$; after this symmetric uncoded cache placement, MAN generates multicast coded messages (by a binary linear network code) that are simultaneously useful to many users; these coded multicast message delivery drastically reduces the download time, or network load, compared to traditional caching strategies. 
The MAN scheme is known to be optimal for shared-link broadcast networks under the constrain of uncoded cache placement 
and optimal to within a factor~two otherwise~\cite{yas2}.

Coded caching has been extended to other network topologies, such as relay networks, 
where cache-aided users communicate with the server through intermediate relays. 
Due to the difficulty of analyzing general relay networks, in this paper we consider the symmetric network known as {\it combination network}~\cite{cachingincom}.
In a $(\Hsf,\rsf,\Msf,\Nsf)$ combination network, 
a server with $\Nsf$ files 
is connected to $\Hsf$ relays (without caches) through $\Hsf$ orthogonal links, and each of the $\Ksf = \binom{\Hsf}{\rsf}$ users (with caches of size $\Msf$ files) is connected to a different subset of the $\rsf$ relays through $\rsf$ orthogonal links.
The goal is to design a caching (placement and delivery) scheme so as to minimize the maximum number of transmitted files among all the links for the worst set of user demands.

\paragraph*{\textbf{Related Work}}  
Two achievable schemes were proposed in~\cite{cachingincom} for combination networks: 
one based on uncoded placement and routing in the delivery phase, and 
the other on MAN placement and a linear code for delivery.
In~\cite{Zewail2017codedcaching}, a scheme was proposed 
based on coded placement that effectively splits the combination network into $\Hsf$ parallel shared-link networks,
each of which serves $\frac{\Ksf\rsf}{\Hsf} = \binom{\Hsf-1}{\rsf-1}$ users by using the shared-link MAN scheme.
Both works benchmarked the performance of their schemes by using a cut-set outer bound as in~\cite{dvbt2fundamental}.


Recently, we have proposed several outer and inner bounds 
for combination networks.
In~\cite{novelwan2017}  we derived novel outer bounds under the constraint of uncoded placement; in particular, we gave various ways to deal with cycles when the `acyclic index coding outer bound' originally proposed in~\cite{ontheoptimality} for shared-link networks is not applicable.
In~\cite{novelwan2017} we proposed two achievable schemes, both with MAN placement and MAN multicast message generation; one works for general relay networks; the other one, for combination networks with $\Msf=\Nsf/\Ksf$, leverages the symmetric topology of combination networks and proposes to further network code the MAN multicast messages so as to enable interference elimination.
%
In~\cite{wan2017novelmulticase}, we introduced a scheme that 
uses MAN placement but with a novel way to generate and deliver multicast messages.
Numerical results show that the combination of our schemes in~\cite{novelwan2017,wan2017novelmulticase} improves on~\cite{Zewail2017codedcaching}. 


\paragraph*{\textbf{Contributions of Paper Organization}}
Section~\ref{sec:model} presents the system model.
Section~\ref{sec:novel scheme} gives the details of the novel proposed scheme, which--following the so called separation approach~\cite{Naderializadeh2017onthoptimality}--uses MAN placement and MAN multicast message generation (both oblivious of the network topology). 
The main contribution of this paper is a novel delivery scheme 
that leverages the symmetries in the network topology and is order optimal for small $\Msf$.
We proposed to deliver the MAN multicast messages 
with the following two-phase scheme:
\\
{\it First phase:} we directly transmit each MAN multicast message to some relays, which forward them to their connected users; such messages are simultaneously useful for $t:=\Ksf\Msf/\Nsf$ users and
will be used as `side information' in the next phase; in this phase, the multicasting gain is $t+1$.
\\ 
{\it Second phase:} not all users are able to decode all their desired file at the end of the first phase;
we thus deliver the MAN multicast messages through a carefully design network code, to let the unsatisfied users recover their demanded file; in this phase, the multicasting gain is $\rsf$.
\\
{\it Performance:} The final network load 
compounds the gain from the MAN multicast message generation and from the novel delivery.  
After balancing the transmission load across the relays, the final load/download time is proportional to $1/\Hsf$.
\\
{\it Order Optimality:}
In Section~\ref{sec:opt}, the proposed scheme is proved to be optimal to within a factor $\fsf:=1+t/\rsf = 1 + \frac{\Ksf\Msf}{\Nsf\rsf}$ under the constraint of uncoded placement.
We remark that for the small cache size regime given by $t=\Ksf\Msf/\Nsf\leq \rsf$, our scheme  is optimal within factor $\fsf \leq 2$ under the constraint of uncoded placement, and--owing to the result in~\cite{yas2}--it is optimal to within a factor $2\fsf \leq 4$ for general placement.
\\
{\it Improvement:}
In Section~\ref{sub:ICICD}, we improve on the proposed scheme by leveraging further multicasting opportunities, and in Section~\ref{sec:numerical} we show numerically that this improved scheme outperforms all known schemes.

\begin{figure}
\centerline{\includegraphics[scale=0.16]{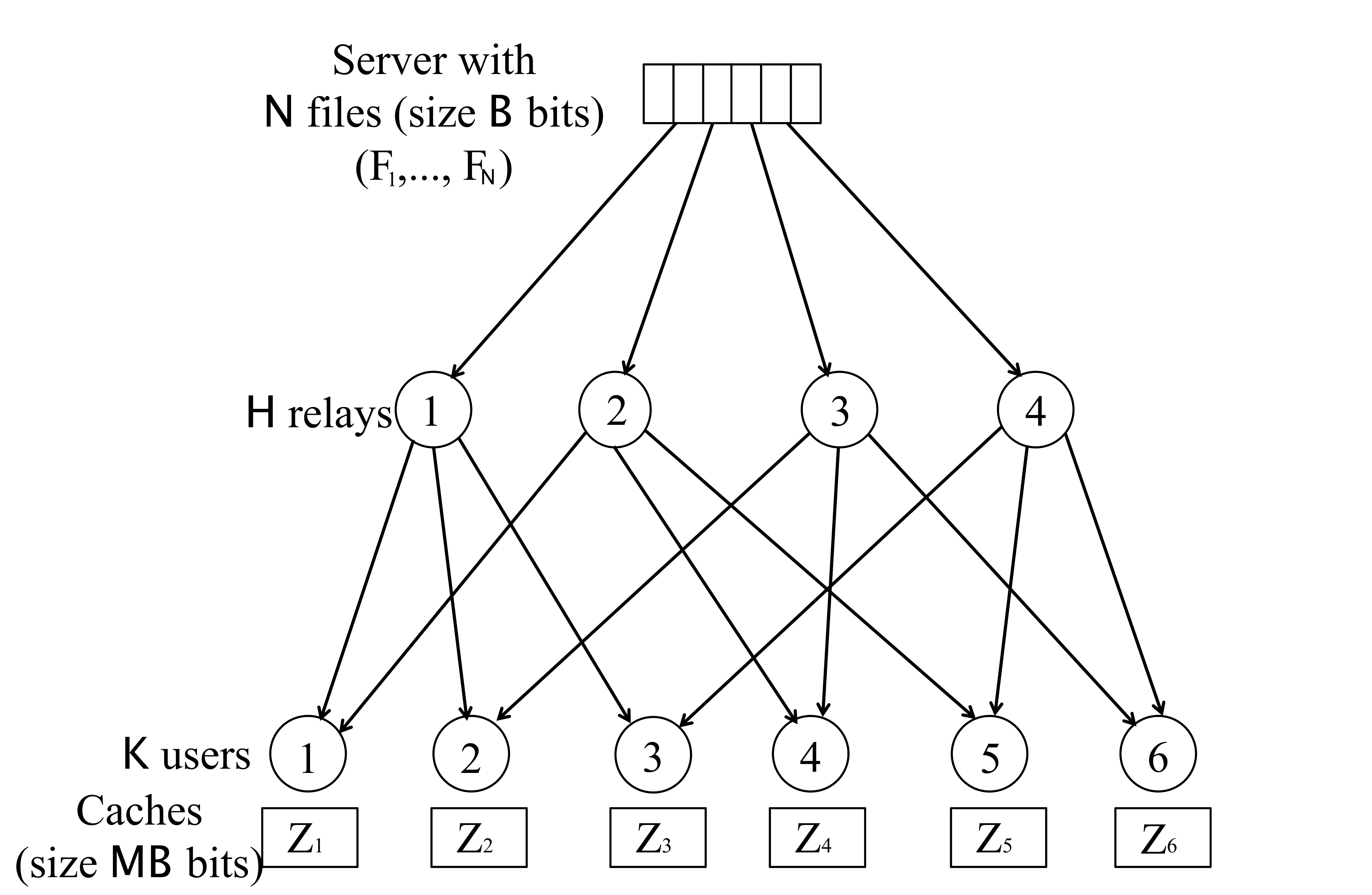}}
\caption{\small A combination network with end-user-caches, with $\Hsf=4$ relays and $\Ksf=6$ users, i.e., $\rsf=2$.}
\label{fig: Combination_Networks}
\vspace{-5mm}
\end{figure}

\section{System Model} 
\label{sec:model}

\paragraph*{\textbf{Notation Convention}}
Calligraphic symbols denote sets, 
bold symbols denote vectors, and 
sans-serif symbols denote system parameters.
We use
$|\cdot|$ to represent the cardinality of a set;
$[a:b]:=\left\{ a,a+1,\ldots,b\right\}$ and $[n] := [1:n]$.
$\mathcal{A\setminus B}:=\left\{ x\in\Ac : x\notin\Bc\right\}$; 
We define the set $\arg \max_{x\in \Xc}f(x) := \big\{x\in \Xc:f(x)=\max_{x\in\Xc}f(x)\big\}$.
$\text{RLC}(m,\Sc)$ represents $m$ random linear combinations of the equal-length packets indexed by $\Sc$;  $m$ random linear combinations of $|\Sc|$ packets are linearly independent with high probability if operations are done on a large enough finite field;
the same can be obtained by using the parity-check matrix of an $(|\Sc|, |\Sc|-m)$ MDS (Maximum Distance Separable) code~\cite{ourisitinnerbound}.

\paragraph*{\textbf{System Model}}
A server has access to $\Nsf$ files denoted by $\{F_1, \cdots, F_\Nsf\}$, each composed of $\Bsf$ i.i.d. bits, and 
is connected to $\Hsf$ relays through $\Hsf$ error-free and interference-free links. The relays are connected to $\Ksf = \binom{\Hsf}{\rsf}$ users nodes 
through $\rsf \, \Ksf$ error-free and interference-free links. 
%
The set of users connected to the $h$-th relay is denoted by $\Uc_{h}, \ h\in[\Hsf]$. 
The set of relays connected to $k$-th user is denoted by $\Hc_{k}, k\in[\Ksf]$.
For each set of users $\Wc\subseteq [\Ksf]$, let $\Hc_{\Wc}=\cup_{k\in \Wc}\Hc_{k}$.
For the combination network in~Fig.~\ref{fig: Combination_Networks}, we have, for example, 
$\Uc_{1}=\{1,2,3\}$,
$\Hc_{1}=\{1,2\}$, and 
$\Hc_{\{1,2\}}=\{1,2,3\}$.

In the placement phase, user $k\in[\Ksf]$ stores information about the $\Nsf$ files in its cache of size $\mathsf{MB}$ bits, where $\Msf \in[0,\Nsf]$.  
We denote the content in the cache of user $k\in[\Ksf]$ by $Z_{k}$ and let $\Zm:=(Z_{1},\ldots,Z_{\Ksf})$.
During the delivery phase, user $k\in[\Ksf]$ demands file $d_{k}\in[\Nsf]$;
the demand vector $\dv:=(d_{1},\ldots,d_{\Ksf})$ is revealed to all nodes. 
Given $(\dv,\Zm)$, the server sends a message $X_{h}$ 
of $\Bsf \, \Rsf_{h}(\dv,\Zm)$ bits to relay $h\in [\Hsf]$. 
Then, relay $h\in [\Hsf]$ transmits a message $X_{h\to k}$ 
of $\Bsf \, \Rsf_{h\to k}(\dv,\Zm)$ bits to user $k \in \Uc_h$. 
User $k\in[\Ksf]$ must recover its desired file $F_{d_{k}}$ from $Z_{k}$ and $(X_{h\to k} : h\in \Hc_k)$ with high probability for some $\Bsf$. 
The objective is to determine the optimal {\it max-link load} 
\begin{align}
\Rsf^{\star}(\Msf)
:=
\min_{\substack{\Zm}} \negmedspace
\max_{\substack{k\in\Uc_h, h\in[\Hsf],\\ \dv\in[\Nsf]^{\Ksf}}} \negmedspace
\max 
\left\{
\Rsf_h(\dv,\Zm),
\Rsf_{h\to k}(\dv,\Zm)
\right\}.
\end{align}
The placement phase is said to be uncoded if each user directly copies some bits  into its cache. 
The max-link load under the constraint of uncoded cache placement is denoted by $\Rsf^{\star}_{\mathrm{u}}(\Msf)$. 
In general, $\Rsf^{\star}_{\mathrm{u}}(\Msf) \geq \Rsf^{\star}(\Msf)$.

\section{Novel Delivery Schemes }
\label{sec:novel scheme}

\subsection{High Level Description of the Proposed Scheme}
\label{sub:CICDhigh}
Our proposed scheme uses MAN placement and MAN multicast message generation, and then it  delivers the MAN multicast messages in two phases.
The scheme is as follows.

\paragraph{MAN placement}
Let $\Msf=t\frac{\Nsf}{\Ksf}$ for some $t\in[0:\Ksf]$. 
Let $F_{i}=(F_{i,\Wc}:\Wc\subseteq [\Ksf],|\Wc|=t)$ be 
a partition of file $F_i, i\in [\Nsf],$ into $\binom{\Ksf}{t}$ equal size sub-files.
User $k\in [\Ksf]$ stores $F_{i,\Wc}$ in its cache if $k\in\Wc$.

\paragraph{MAN multicast message generation} 
Let the demanded vector be $\dv$.
For each $\Jc\subseteq [\Ksf]$ of cardinality $|\Jc|=t+1$, 
multicast messages are generated as follows
\begin{align}
W_{\Jc} := \oplus_{j\in\Jc}F_{d_{j},\Jc\setminus\{j\}}.
\label{eq:MAN multicast messages}
\end{align}
User $k\in\Jc$ can recover $F_{d_{k},\Jc\backslash\{k\}}$ from $W_{\Jc}$ and the cached sub-files
becuse it knows $F_{d_{j},\Jc\backslash\{j\}}$ 
for all $j\in\Jc$ where $j\neq k$.

\paragraph{Novel delivery} 
At a high level, our proposed delivery works as follows.
We directly transmit each  MAN message to some relays in the first phase such that each $W_{\Jc}$ can be recovered by a subset of users in $\Jc$;
these messages can be seen as side information for the second phase. 
In the second phase, we design linear combinations of the MAN messages such that the users in $\Jc$ who did not recover $W_{\Jc}$ previously can do so.
We illustrate this idea by means of an example first.

\subsection{Example~1}
\label{ex:ex1 of CICD}
Consider the network in Fig.~\ref{fig: Combination_Networks} with $\Nsf=\Ksf=6$, $\Msf=t=2$ and let $\mathbf{d}=(1:6)$. For each $\Jc\subseteq[\Ksf]=[6]$ where $|\Jc|=t+1=3$, the MAN multicast messages in~\eqref{eq:MAN multicast messages} 
contain $\Bsf/\binom{\Ksf}{t} =  \Bsf/15$ bits.
Let us now look at the two-phase delivery. 

\paragraph*{First phase} 
For each 
$\Jc\subseteq[\Ksf]=[6]$ of size $|\Jc|=t+1=3$, 
we compute $\Sc_{\Jc} := \arg\max_{h\in [\Hsf]}|\Uc_h\cap \Jc|$ (i.e.,
the set of relays each of which is connected to the largest number of users  in $\Jc$),
then partition $W_{\Jc}$ into $|\Sc_{\Jc}|$ equal-length parts, denoted as $W_{\Jc}=\{W^{|\Sc_{\Jc}|}_{\Jc,h}:h\in \Sc_{\Jc}\}$, and finally transmit $W^{|\Sc_{\Jc}|}_{\Jc,h}$ to relay $h\in \Sc_{\Jc}$.

For example, 
for $\Jc=\{1,2,3\}$,
relay~$1$ is connected to three users in $\Jc$ (user $1$, $2$ and $3$), 
relay $2$ is connected to one user in $\Jc$ (user $1$), 
relay $3$ is connected to one user in $\Jc$ (user $2$), and 
relay $4$ is connected to one user in $\Jc$ (user $3$). 
So we have $\Sc_{\{1,2,3\}} = \arg\max_{h\in[4]}|\Uc_h\cap \{1,2,3\}|=\{1\}$.
Therefore, we have $W_{\{1,2,3\}}=\{W^1_{\{1,2,3\},1}\}$.

Similarly, for $\Jc=\{1,2,4\}$,
relay~$1$ is connected to two users in $\Jc$ (users  $1$ and $2$), 
relay $2$ is connected to two users in $\Jc$ (users   $1$ and $4$), 
relay $3$ is connected to two users in $\Jc$ (users   $2$ and $4$), and 
relay $4$ is not connected to any users in $\Jc$. 
So we have $\Sc_{\Jc} =\arg\max_{h\in[4]}|\Uc_h\cap \{1,2,4\}|=\{1,2,3\}$.
Therefore, we have $W_{\{1,2,4\}}=\{W^3_{\{1,2,4\},1}, W^3_{\{1,2,4\},2}, W^3_{\{1,2,4\},3}\}$.

By considering all the subsets $\Jc$, 
the server has sent to relay~$1$ (and similarly for all other relays)
\begin{align*}
&W^1_{\{1,2,3\},1}, W^3_{\{1,2,4\},1} ,W^2_{\{1,2,5\},1}, W^2_{\{1,2,6\},1}, W^2_{\{1,3,4\},1}, 
\\
&W^3_{\{1,3,5\},1}, W^2_{\{1,3,6\},1}, W^2_{\{2,3,4\},1}, W^2_{\{2,3,5\},1}, W^3_{\{2,3,6\},1},
\end{align*}
for a total of $\Bsf/15+6\Bsf/30+3\Bsf/45=\Bsf/3$ bits;
these messages are then forwarded by relay~$1$ to the users in $\Uc_1$.

\paragraph*{Second phase} 
Let us first focus on message $W_{\{1,2,3\}}=\{W^1_{\{1,2,3\},1}\}$. 
From the first phase, $W^1_{\{1,2,3\},1}$ can be recovered by the users in  $\Uc_1 \cap \{1,2,3\} = \{1,2,3\}$
from the transmission by relay~$1$. 
Hence, $W^1_{\{1,2,3\},1}$ need not to be transmitted in the second phase as all users that must recover it have done so.

For $W_{\{1,2,4\}}=\{W^3_{\{1,2,4\},1},W^3_{\{1,2,4\},2},W^3_{\{1,2,4\},3}\}$, from the first phase, $W^3_{\{1,2,4\},1}$ can be recovered by the users in  $\Uc_1 \cap \{1,2,4\} = \{1,2\}$
but not by the users in  $\Uc_1 \backslash \{1,2,4\} = \{4\}$
from the transmission by relay~$1$.
%
In the second phase, we thus aim to transmit $W^3_{\{1,2,4\},1}$ to user $4$. 
We divide $W^3_{\{1,2,4\},1}$ into $\rsf=2$ non-overlapping and equal-length pieces, $W^3_{\{1,2,4\},1}=\{W^3_{\{1,2,4\},1,h}:h\in \Hc_4\}$, with $\Hc_4=\{2,3\}$. We then let user $4$ recover $W^3_{\{1,2,4\},1,2}$ from relay $2$, and $W^3_{\{1,2,4\},1,3}$ from relay $3$. In order to do so, since user $1$, who is also connected to relay $2$, knows $W^3_{\{1,2,4\},1,2}$, we put $W^3_{\{1,2,4\},1,2}$ into $\Pc^{2}_{4,\{1\}}$;
here $\Pc^{2}_{4,\{1\}}$ represents the set of bits needed to be recovered by user $4$ (the first entry in the subscript) from relay $2$ (the superscript) and already known by user $1$ (the second entry in the subscript) who is also connected to relay $2$ (the superscript). Similarly, we put $W^3_{\{1,2,4\},1,3}$ in $\Pc^{3}_{4,\{2\}}$. After considering all the pieces of the multicast messages $W_{\Jc}$ 
that need transmission in the second phase for relay~$1$ we have (similarly for all other relays) 
\begin{align*}
&\Pc^{1}_{1,\{2\}}=\{W^3_{\{1,2,4\},3,1},W^2_{\{1,2,6\},3,1},W^2_{\{1,4,6\},3,1}\},\\ 
&\Pc^{1}_{1,\{3\}}=\{W^3_{\{1,3,5\},4,1},W^2_{\{1,3,6\},4,1},W^2_{\{1,5,6\},4,1}\}, \\
&\Pc^{1}_{2,\{1\}}=\{W^3_{\{1,2,4\},2,1},W^2_{\{1,2,5\},2,1},W^2_{\{2,4,5\},2,1}\},\\
&\Pc^{1}_{2,\{3\}}=\{W^2_{\{2,3,5\},4,1},W^3_{\{2,3,6\},4,1},W^2_{\{2,5,6\},4,1}\},\\
&\Pc^{1}_{3,\{1\}}=\{W^2_{\{1,3,4\},2,1},W^3_{\{1,3,5\},2,1},W^2_{\{3,4,5\},2,1}\},\\ 
&\Pc^{1}_{3,\{2\}}=\{W^2_{\{2,3,4\},3,1},W^3_{\{2,3,6\},3,1},W^2_{\{3,4,6\},3,1}\};
\end{align*}
each $\Pc$'s contains $\Bsf/90+2\Bsf/60=2\Bsf/45$ bits.
We transmit 
\begin{align*}
\Pc^{1}_{1,\{2\}}\oplus\Pc^{1}_{2,\{1\}} := 
\{
W^3_{\{1,2,4\},3,1}\oplus W^3_{\{1,3,5\},4,1}, \\
W^2_{\{1,2,6\},3,1}\oplus W^2_{\{1,3,6\},4,1}, \
W^2_{\{1,4,6\},3,1}\oplus W^2_{\{1,5,6\},4,1}
\}
\end{align*}
to relay~$1$, such that user $1$ knowing $\Pc^{1}_{2,\{1\}}$ can recover $\Pc^{1}_{1,\{2\}}$, and user $2$ knowing $\Pc^{1}_{1,\{2\}}$ can recover $\Pc^{1}_{2,\{1\}}$. 
Similarly, the server transmits $\Pc^{1}_{1,\{3\}}\oplus\Pc^{1}_{3,\{1\}}$ and $\Pc^{1}_{2,\{3\}}\oplus\Pc^{1}_{3,\{2\}}$ to relay~$1$. 
At the end of this phase, for each $\Vc\subseteq \Uc_1$, relay~$1$ (and similarly for the other relays) forwards $\underset{k\in\Jc}{\oplus}\Pc^{1}_{k,\Vc\setminus \{k\}}$ to the users in $\Vc$. 
The total number of transmitted bits from the server to a relay in this phase is $2\Bsf/15$.

In conclusion, the  achieved max link-load  is $1/3+2/45=7/15$, while the max link-loads of the schemes in~\cite{novelwan2017},~\cite{cachingincom} and~\cite{Zewail2017codedcaching} are $8/15$, $2/3$ and $1/2$, respectively. 
The outer bound with uncoded placement in \cite[Thm.4]{novelwan2017} is $7/17$.

\subsection{Detailed Description of the Proposed Scheme}
\label{sub:CICD}

\paragraph*{First phase} 
For each $W_{\Jc}$ in~\eqref{eq:MAN multicast messages} where $\Jc\subseteq [\Ksf]$ and $|\Jc|=t+1$, we find
$\Sc_{\Jc} := \arg\max_{h\in[\Hsf]}|\Uc_h\cap\Jc|$ (i.e., the set of relays 
each relay in which is connected to the largest number of users in $\Jc$).  We then partition $W_{\Jc}$ into $|\Sc_{\Jc}|$ non-overlapping equal-length pieces and denote $W_{\Jc}=\{W^{|\Sc_{\Jc}|}_{\Jc,h}:h\in\Sc_{\Jc}\}$.  The server transmits  $W^{|\Sc_{\Jc}|}_{\Jc,h}$ to relay $h\in \Sc_{\Jc}$, and relay $h\in \Sc$ transmits $W^{|\Sc_{\Jc}|}_{\Jc,h}$ to the users in $\Uc_h$. 

\paragraph*{Second phase} 
For each $W_{\Jc}$ as in the first phase
the users in $\Jc\cap \Uc_h, h\in\Sc_{\Jc},$ can recover $W^{|\Sc_{\Jc}|}_{\Jc,h}$; 
thus the second phase aims to transmit $W^{|\Sc_{\Jc}|}_{\Jc,h}$ to the users in $\Jc\setminus \Uc_h$. 
For each user $k\in\Jc\setminus \Uc_h$,  $W^{|\Sc_{\Jc}|}_{\Jc,h}$ is divided into $\rsf$ non-overlapping and equal-length pieces and denoted as $W^{|\Sc_{\Jc}|}_{\Jc,h}=\{W^{|\Sc_{\Jc}|}_{\Jc,h,h^{\prime}}:h^{\prime}\in \Hc_{k}\}$.
We aim to let user $k\in \Jc\setminus \Uc_h$  recover  $W^{|\Sc_{\Jc}|}_{\Jc,h,h^{\prime}}$ from relay $h^{\prime}\in \Hc_{k}$.   
For relays $h,h^{\prime}$ and user $k$, where user $k$ is connected to relay $h^{\prime}$ but not to relay $h$,  we define
\begin{align}
\Qc^{k}_{h,h^{\prime}}:=\big\{j\in \Uc_{h}\cap \Uc_{h^{\prime}}:\Hc_{j}\subseteq \Hc_{k}\cup \{h\}\big\}
\label{eq:notpretty}
\end{align}
and put $W^{|\Sc_{\Jc}|}_{\Jc,h,h^{\prime}}$ in $\Pc^{h^{\prime}}_{k,\Qc^{k}_{h,h^{\prime}}}$ representing the set of bits known by the users in $\Qc^{k}_{h,h^{\prime}}$ and to  be recovered  by user $k$ from relay $h^{\prime}$. Note $|\Qc^{k}_{h,h^{\prime}}|=\rsf-1$, as explained in Remark~\ref{footnote:Qc}. 
Finally, for each relay $h\in[\Hsf]$ and each set $\Vc\subseteq\Uc_h$ where $|\Vc|=\rsf$, the server transmits $\oplus_{k\in\Vc}\Pc^{h}_{k,\Jc\setminus \{k\}}$ to relay $h$, which forwards it to the users in $\Vc$.

\begin{rem}
\label{footnote:Qc}
The two partition steps for each $W_{\Jc}$ (e.g., $W^{|\Sc_{\Jc}|}_{\Jc,h}$ and  $W^{|\Sc_{\Jc}|}_{\Jc,h,h^{\prime}}$) ensure that the number of bits transmitted from the server to each relay is the same.
 So the achieved max link-load is  proportional to $1/\Hsf$.

We put $W^{|\Sc_{\Jc}|}_{\Jc,h,h^{\prime}}$ in $\Pc^{h^{\prime}}_{k,\Qc^{k}_{h,h^{\prime}}}$ where $\Qc^{k}_{h,h^{\prime}}$ is defined in~\eqref{eq:notpretty};
among the relays $h,h^{\prime}$ user $k$ is only connected to relay $h^{\prime}$.  Since the users in $\Qc^{k}_{h,h^{\prime}}$ are connected to relays $h$ and $h^{\prime}$ simultaneously and the connected relays of these users are in the set $\Hc_{k}\cup \{h\}$ including $\rsf+1$ relays, 
one has $|\Qc^{k}_{h,h^{\prime}}|=\binom{\rsf+1-2}{\rsf-2}=\rsf-1$. 
By the symmetry of the combination network, for each relay $a\in \Hc_{k}\setminus \{h^{\prime}\}$,  there must exist one set $\Jc^{\prime}$ with $|\Jc^{\prime}|=t$ where
$a$ is in the set $\arg\max_{b\in[\Hsf]}|\Uc_{b}\cap\Jc^{\prime}|$, which also includes $|\Sc_{\Jc}|$ elements, and the user (assumed to be $k^{\prime}$) connected to relays in $(\Hc_k\cup \{h\})\setminus \{a\}$ is also in $\Jc^{\prime}$.   Since user $k^{\prime}$ is connected to relays $h$ and $h^{\prime}$, one has  $k^{\prime}\in \Qc^{k}_{h,h^{\prime}}$.
In addition, user $k^{\prime}$ needs to recover $W^{|\Sc_{\Jc^{\prime}}|}_{\Jc^{\prime},a,h^{\prime}}$ from relay $h^{\prime}$, whose length is equal to the length of $W^{|\Sc_{\Jc}|}_{\Jc,h,h^{\prime}}$. 
Notice that   $W^{|\Sc_{\Jc^{\prime}}|}_{\Jc^{\prime},a}$ is directly transmitted to relay $a$, so $W^{|\Sc_{\Jc^{\prime}}|}_{\Jc^{\prime},a,h^{\prime}}$ is
known by the $\rsf-2$ users in $\Qc^{k}_{h,h^{\prime}} \setminus \{k^{\prime}\}$ and by user $k$. So we can add $W^{|\Sc|}_{\Jc,h,h^{\prime}}$ and $W^{|\Sc_{\Jc^{\prime}}|}_{\Jc^{\prime},a,h^{\prime}}$ such that user $k$ and $k^{\prime}$ can recover their desired pieces. Similarly, there are $|\Hc_{k}\setminus \{h^{\prime}\}|=\rsf-1$ relays as relay $a$. So $W^{|\Sc|}_{\Jc,h,h^{\prime}}$ can be added with the other $\rsf-1$ pieces with the same length  (each of which is demanded by one user in $\Qc^{k}_{h,h^{\prime}}$) and then be transmitted to relay $h^{\prime}$.
\end{rem}

\paragraph*{Performance}
For each $W_{\Jc}$ 
the server directly transmits $W^{|\Sc_{\Jc}|}_{\Jc,h}$ to relay $h\in\Sc_{\Jc}$ in the first phase  for a total of $|W_{\Jc}| = \Bsf/\binom{\Ksf}{t}$ bits. 
In the second phase, for relay $h\in\Sc_{\Jc}$, $|\Jc\setminus \Uc_h|$ users recover $W^{|\Sc_{\Jc}|}_{\Jc,h}$. So
the server transmits $W^{|\Sc_{\Jc}|}_{\Jc,h,h^{\prime}}$ to each user $k\in \Jc\setminus \Uc_h$ for $h^{\prime}\in \Hc_k$) in one linear combination with other $\rsf-1$ pieces of the same length (equal to $\frac{\Bsf}{\rsf\binom{\Ksf}{t}|\Sc_{\Jc}|}$. Hence, the total link-load to transmit $W_{\Jc}$ is $\frac{1}{\binom{\Ksf}{t}}+\frac{|\Sc_{\Jc}||\Jc\setminus \Uc_h|}{\rsf\binom{\Ksf}{t}|\Sc_{\Jc}|}=\frac{1+|\Jc\setminus \Uc_h|/\rsf}{\binom{\Ksf}{t}}$.
By the symmetry of network, the number of transmitted bits to each relay is the same and given in the following theorem.
\begin{thm}
\label{thm: load of CICD}
For a $(\Hsf,\rsf,\Msf,\Nsf)$ combination network
with $t=\Ksf\Msf/\Nsf\in [0:\Ksf]$,  the max link-load is
\begin{align}
&\Rsf^{\star}_{\mathrm{u}}\leq \Rsf_{1} := 
\sum_{\Jc\subseteq[\Ksf]:|\Jc|=t}\frac{1+\min_{h\in[\Hsf]}|\Jc\setminus \Uc_h|/\rsf}{\Hsf\binom{\Ksf}{t}},
\label{eq:load of CICD}
\end{align}
where $\Rsf_1$ is achieved by the scheme in Section~\ref{sub:CICD}.  The tradeoff between memory size and max link-load is the lower convex envelope of the above points.
\end{thm}

\subsection{Optimality Results for the Proposed Scheme}
\label{sec:opt}
\begin{thm}
\label{thm:optimality of CICD}
For a $(\Hsf,\rsf,\Msf,\Nsf)$ combination network with
$\Nsf\geq \Ksf=\binom{\Hsf}{\rsf}$ and $\Msf=\frac{t\Nsf}{\Ksf}, \ t\in[0:\Ksf]$, the scheme in Section~\ref{sub:CICD} is optimal within a factor $1+t/\rsf$ under the constraint of uncoded cache placement, 
and to within a factor of $2(1+t/\rsf)$ otherwise.
\end{thm}

\begin{IEEEproof}
We compare Theorem~\ref{thm:optimality of CICD} to the cut-set outer bound in~\cite[Eq.(12)]{novelwan2017}.  For each $\Jc\subseteq [\Ksf]$ where $|\Jc|=t+1$, there exist some relays connected to at least one user in $\Jc$, and thus one has $\min_{h\in[\Hsf]}|\Jc\setminus \Uc_h|\leq |\Jc|-1=t$. 
Hence, from~\eqref{eq:load of CICD}, 
\begin{align}
\Rsf_{1}\leq \binom{\Ksf}{t+1}\frac{1+t/\rsf}{\Hsf\binom{\Ksf}{t}}.
\label{eq:outer bound of CICD}
\end{align}
From~\cite{novelwan2017}, it was proved that $\Rsf^{\star}_{\mathrm{u}}$ is lower  bounded by the convex hull of $\Big(\frac{\Nsf t}{\Ksf},\frac{\binom{\Ksf}{t+1}}{\Hsf\binom{\Ksf}{t}}\Big)$ for $t\in [0:\Ksf]$. Hence, the scheme in Section~\ref{sub:CICD} is order optimal within a factor of $1+t/\rsf$ under the constraint of uncoded cache placement. 
Since the multiplicative gap between uncoded and coded placements in shared-link network (which is used in \cite{novelwan2017} to derive the converse bound) is not larger than $2$~\cite{yas2},  the scheme in Section~\ref{sub:CICD} is order optimal within a factor of $2(1+t/\rsf)$.
\end{IEEEproof}

\begin{rem}
\label{rem:discussion of optimality CICD}
Under the assumptions of Theorem~\ref{thm:optimality of CICD}, we have:
(a) if $t \leq \rsf$  (i.e., small cache size regime), the scheme in Section~\ref{sub:CICD} is order optimal within a constant factor no larger than $(1+t/\rsf)\leq 2$, and
(b) if $t/\rsf \to 0$ the scheme in Section~\ref{sub:CICD} is optimal under under uncoded placement.
The schemes in~\cite{cachingincom,novelwan2017,Zewail2017codedcaching} are order optimal, under the uncoded cache placement, to within a factor $\Hsf/\rsf$,  which can be arbitrary large. 
\end{rem}

The scheme in Section~\ref{sub:CICD} can be improved by leveraging  multicasting opportunities 
as the following example shows. 

\subsection{Example~2}
\label{ex:ex2 of ICICD}

Consider the network in Fig.~\ref{fig: Combination_Networks} with $\Nsf=\Ksf=6$, $\Msf=t=3$ and let $\mathbf{d}=(1:6)$.  For each $\Jc\subseteq[\Ksf]=[6]$ where $|\Jc|=t+1=4$, 
each  MAN multicast message in~\eqref{eq:MAN multicast messages} 
contains $\Bsf/\binom{\Ksf}{t}=\Bsf/20$ bits.  We proceed now to describe the improved delivery compared to the scheme in Section~\ref{sub:CICD}.

\paragraph*{First phase} 
This step is the same as in Section~\ref{sub:CICD} with the exception that
each coded messages $W_{\Jc}$ 
is divided into $\Bsf/P$ packets for some large enough length $P$ (possible since $\Bsf$ can be taken arbitrary large).

\paragraph*{Second phase} 
If one piece of multicast coded message needs to be recovered by a user in the second phase, the scheme in Section~\ref{sub:CICD} divides this piece into $\rsf$ non-overlapping parts and let the user recover one
 different part from each of its connected relays. 
For example, in the first phase $W_{\{1,2,5,6\}}$ is divided into $|\Sc_{\{1,2,5,6\}}|=4$ non-overlapping and equal-length pieces, $W_{\{1,2,5,6\}}=\{W^4_{\{1,2,5,6\},h}:h\in [4]\}$. We directly transmit $W^{4}_{\{1,2,5,6\},4}$ to relay $4$, which forwards it to users $3$, $5$ and $6$. So in the second phase, users $1$ and $2$ must recover $W^{4}_{\{1,2,5,6\},4}$. 
The scheme in Section~\ref{sub:CICD} treats $W^4_{\{1,2,5,6\},4}$ demanded by user $1$ and $2$ as two independent pieces. It lets user $1$ recover $|W^4_{\{1,2,5,6\},4}|/2$ bits from relay~$1$ and $|W^4_{\{1,2,5,6\},4}|/2$ bits from relay $2$ in two linear combinations, and lets user $2$ recover $|W^4_{\{1,2,5,6\},4}|/2$ bits from relay~$1$ and $|W^4_{\{1,2,5,6\},4}|/2$ bits from relay $3$ in two other  ones.
Transmitting one message `twice' is  inefficient.

Instead, we can leverage the following multicasting opportunity.
We put $\text{RLC}(|W^4_{\{1,2,5,6\},4}|/(2P),W^4_{\{1,2,5,6\},4})$ in $\Xc^{1}_{\{1,2\},\{3\}}$, where  $\Xc^{1}_{\{1,2\},\{3\}}$ is the set of packets needed to be recovered by users in $\{1,2\}$ (first part of the subscript) from relay~$1$ (superscript) and already known by the users in $\{3\}$ (second part of the subscript) who are also connected to relay~$1$ (superscript). The number of packets in $\Xc^{1}_{\{1,2\},\{3\}}$ is $|\Xc^{1}_{\{1,2\},\{3\}}|/P$.
We then encode the messages at relay~$1$ as
\begin{align*}
\Xc^{1}_{\{1\},\{2\}}\oplus\Xc^{1}_{\{2\},\{1\}}, \
\Xc^{1}_{\{1\},\{3\}}\oplus\Xc^{1}_{\{3\},\{1\}}, \
\Xc^{1}_{\{2\},\{3\}}\oplus\Xc^{1}_{\{3\},\{2\}},
\end{align*}
where we used the same convention as before when it comes to `summing' sets.
We also send $\text{RLC}(2|\Xc^{1}_{\{2,3\},\{1\}}|/P,\Xc^{1}_{\{1,2\},\{3\}}\cup\Xc^{1}_{\{1,3\},\{2\}}\cup\Xc^{1}_{\{2,3\},\{1\}})$ to relay~$1$. 
Note that the users in $\{1,2,3\}$ know $|\Xc^{1}_{\{2,3\},\{1\}}|/P$ packets of $\Xc^{1}_{\{1,2\},\{3\}}\cup\Xc^{1}_{\{1,3\},\{2\}}\cup\Xc^{1}_{\{2,3\},\{1\}}$. So if the server transmits  $2|\Xc^{1}_{\{2,3\},\{1\}}|/P$ random linear combinations of those packets to relay~$1$, which will then forward them to its connected users, each user can recover all of the packets of $\Xc^{1}_{\{1,2\},\{3\}}\cup\Xc^{1}_{\{1,3\},\{2\}}\cup\Xc^{1}_{\{2,3\},\{1\}}$ with high probability provided that $\Bsf\to \infty$. 

The  max link-load of these improved two-phases scheme is $\left.\frac{15}{4\binom{K}{t}}+\frac{17}{8\binom{K}{t}}\right|_{\Ksf=6,t=3}=0.29375$, which is less than the max link-load  of the scheme in Section~\ref{sub:CICD} (equal to $0.3$); for the same set of parameters, the max link-loads of the schemes in~\cite{novelwan2017},~\cite{cachingincom} and~\cite{Zewail2017codedcaching} are $0.375$, $0.375$ and $1/3$, respectively.
The outer bound with uncoded placement in \cite[Thm.4]{novelwan2017} is $0.25$.

\subsection{Detailed Description of the Improved Delivery Scheme}
\label{sub:ICICD}
\paragraph*{First phase} 
This step is the same as in Section~\ref{sub:CICD} with the exception that
each coded messages $W_{\Jc}$ 
is divided into $\Bsf/P$ packets for some large enough length $P$
\paragraph*{Second phase} 
For each  $W_{\Jc}$ where $\Jc\subseteq [\Ksf]$ and $|\Jc|=t+1$, and each $h\in\Sc_{\Jc}$,  the second phase is used to transmit $W^{|\Sc_{\Jc}|}_{\Jc,h}$ to the users in $\Jc\setminus \Uc_h$. In this paragraph, to simplify the notation, we let $\Ac:=\Uc_{h^{\prime}}\cap (\Jc\setminus \Uc_h)\neq \emptyset$ and $\Bc:=\big\{j\in \Uc_{h}\cap \Uc_{h^{\prime}}:\Hc_{j}\subseteq \Hc_{\Ac}\cup \{h\}\big\}$. For each $h^{\prime}\in [\Hsf]\setminus \{h\}$ where $\Ac\neq \emptyset$, we add $|W^{|\Sc_{\Jc}|}_{\Jc,h}|/(\rsf P)$ random linear combinations of packets of $W^{|\Sc_{\Jc}|}_{\Jc,h}$ in $\Xc^{h^{\prime}}_{\Ac,\Bc}$ representing the packets to be recovered by users in $\Ac$ from relay $h^{\prime}$ and already known by the users in $\Bc$ who are also connected to relay $h^{\prime}$.  

We aim to let each user $k\in \Jc\setminus \Uc_h$  recover all the sets of packets $\Xc^{h^{\prime}}_{\Wc_1,\Wc_2}\neq \emptyset$ where $h^{\prime}\in\Uc_{k}$ and $k\in \Wc_1$, such that he can recover $|W^{|\Sc_{\Jc}|}_{\Jc,h}|/P$ random linear combinations of packets in $W^{|\Sc_{\Jc}|}_{\Jc,h}$ and then recover $W^{|\Sc_{\Jc}|}_{\Jc,h}$ with high probability provided that $\Bsf\to \infty$. We use a two-stage coding.

{\it Stage~1:} For each relay $h\in[\Hsf]$ and each set $\Vc\subseteq \Uc_h$, we encode all the sets of packets $\Xc^{h}_{\Wc_1,\Wc_2}\neq \emptyset$ where $\Wc_1\cup\Wc_2=\Vc$, by $\Lc^{h}_{\Vc}=\text{RLC}(c/P,\Cc)$ where
\begin{align*}
&\Cc=\underset{\Wc_1,\Wc_2:\Wc_1\cup\Wc_2=\Vc}{\cup}\Xc^{h}_{\Wc_1,\Wc_2},\\
&c=\max_{k\in\Vc}\sum_{\Wc_1,\Wc_2:\Wc_1\cup\Wc_2=\Vc,k\notin \Wc_2}|\Xc^{h}_{\Wc_1,\Wc_2}|,
\end{align*}  
where $c/P$ is the maximal number of packets in $\Cc$ not known by the users in $\Vc$.
So from  $\Lc^{h}_{\Vc}$, each user $k\in \Vc$ can recover all the sets $\Xc^{h}_{\Wc_1,\Wc_2}\neq \emptyset$ where $k\in \Wc_1$.

{\it Stage~2:} If for each set $\Xc^{h}_{\Wc_1,\Wc_2}$, where $\Wc_1\cup\Wc_2=\Vc$ and $|\Xc^{h}_{\Wc_1,\Wc_2}|>0$, we have $k\in \Wc_2$, it can be seen that user $k$ already knows $\Lc^{h}_{\Vc}$ from the first stage. Hence, in one relay $h\in [\Hsf]$, we can encode $\Lc^{h}_{\Vc}$ for all $\Vc\subseteq\Uc_h$ where $\Lc^{h}_{\Vc}\neq \emptyset$ by $\text{RLC}(c^{\prime}/P,\Cc^{\prime})$ where 
\begin{align*}
&\Cc^{\prime}=\underset{\Vc\subseteq\Uc_h:\Lc^{h}_{\Vc}\neq \emptyset}{\cup} \Lc^{h}_{\Vc},\\
&c^{\prime}=\max_{k\in \Uc_h}\sum_{\Vc\subseteq\Uc_h:\Lc^{h}_{\Vc} \textrm{is unknown to }k}|\Lc^{h}_{\Vc}|.
\end{align*}
We transmit $\text{RLC}(c^{\prime}/P,\Cc^{\prime})$ to relay $h$ and relay $h$ then forwards $\text{RLC}(c^{\prime}/P,\Cc^{\prime})$ to users in $\Uc_h$.

\begin{figure}
\centerline{\includegraphics[scale=0.6]{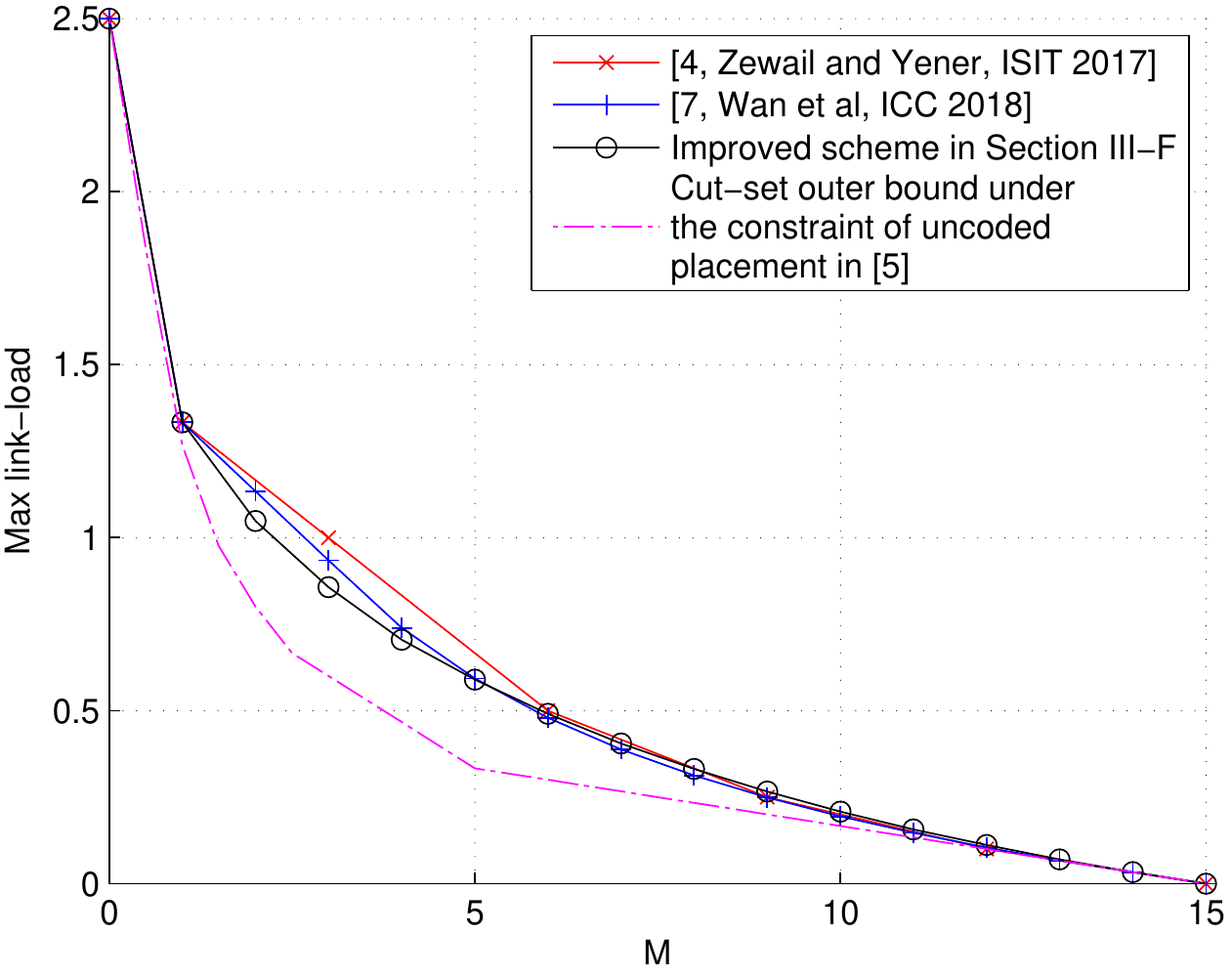}}
\caption{\small A combination network in centralized caching systems with end-user-caches, with $\Hsf=6$ relays,  $\Nsf=\Ksf=\binom{\Hsf}{\rsf}$ and $\rsf=2$.}
\label{fig:rateh6r2}
\vspace{-5mm}
\end{figure}

\section{Numerical Comparisons and Conclusions}
\label{sec:numerical}
In Fig.~\ref{fig:rateh6r2} we compare the performance of the proposed schemes with those available in the literature for the centralized combination network with $\Hsf=6$, $\rsf=2$, and $\Nsf=\Ksf=15$. 
Notice that when $\Msf=\Nsf/\Ksf$, for each illustrated scheme, we use the interference elimination scheme proposed in~\cite{novelwan2017} (which is optimal under uncoded cache placement). We do not plot  the schemes in~\cite{cachingincom} 
 because it performs worse than the others plotted here. Our proposed scheme in Section~\ref{sub:ICICD} outperforms all other schemes when $\Msf$ is not too large (here, $\Msf\leq 5.5$). Compared to the scheme in~\cite{Zewail2017codedcaching} and  our recent results in~\cite{wan2017novelmulticase}, the main advantage of the scheme in Section~\ref{sub:CICD}  is to characterize the order optimality within a constant factor for small cache size. 


\bibliographystyle{IEEEtran}
\bibliography{IEEEabrv,IEEEexample}

\end{document}